\documentclass[
  man,            
  apa7,
  floatsintext,
  donotrepeattitle
]{apa7}

\usepackage[american]{babel}
\usepackage{csquotes}
\usepackage{graphicx}
\usepackage{geometry}
\usepackage{booktabs}
\usepackage{tabularx}
\usepackage{pdflscape}
\usepackage{array}
\usepackage{booktabs}   
\usepackage{tabularx}   
\usepackage{ragged2e}   
\usepackage{rotating}   
\usepackage{textcomp}   

\usepackage{amsmath}
\usepackage{hyperref}
\usepackage{soul, comment}
\usepackage{booktabs}   
\usepackage{array}      
\usepackage{setspace}   
\usepackage{enumitem}
\usepackage{csquotes}
\usepackage[
  style=apa,
  backend=biber,
  sortcites=true,
  sorting=nyt,
  doi=true,
  url=true,
  isbn=false
]{biblatex}

\usepackage{tikz}
\usetikzlibrary{arrows.meta, positioning, shapes, calc}

\usepackage{graphicx}
\usepackage{caption}

\DeclareLanguageMapping{american}{american-apa}

\usepackage{textcomp} 
\newcolumntype{L}{>{\RaggedRight\arraybackslash}X}
\newcommand{\apacomm}{\char44\ }
\title{Boosting Metacognition in Entangled Human--AI Interaction to Navigate Cognitive--Behavioral Drift}
\shorttitle{Boosting Metacognition in Entangled Human--AI Interaction}

\author{
Ezequiel Lopez-Lopez\textsuperscript{1,2},
Christoph M. Abels\textsuperscript{2,3},
Philipp Lorenz-Spreen\textsuperscript{1,2},
Stephan Lewandowsky\textsuperscript{3,4},
Stefan M. Herzog\textsuperscript{2} \vspace{2cm}
}
\authorsaffiliations{
\textsuperscript{1}Center Synergy of Systems\apacomm Technische Universität Dresden\apacomm Germany\\
\textsuperscript{2}Center for Adaptive Rationality\apacomm Max Planck Institute for Human Development\apacomm Germany\\
\textsuperscript{3}Department of Psychology\apacomm University of Potsdam\apacomm Germany\\
\textsuperscript{4}School of Psychological Science\apacomm University of Bristol\apacomm United Kingdom
}

\authornote{
  Ezequiel Lopez-Lopez ORCID: \href{https://orcid.org/0000-0001-8709-8044}{0000-0001-8709-8044}
  
  Christoph M. Abels ORCID: \href{https://orcid.org/0000-0003-2012-5372}{0000-0003-2012-5372}
  
  Philipp Lorenz-Spreen ORCID: \href{https://orcid.org/0000-0001-6319-4154}{0000-0001-6319-4154}
  
  Stephan Lewandowsky ORCID: \href{https://orcid.org/0000-0003-1655-2013}{0000-0003-1655-2013}
  
  Stefan M. Herzog ORCID: \href{https://orcid.org/0000-0003-2329-6433}{0000-0003-2329-6433}

\vspace{2cm}

E.L-L. and S.M.H. acknowledge funding by the Deutsche Forschungsgemeinschaft (DFG), project number 458366841 (POLTOOLS: Assisting behavioral science and evidence-based policy making using online machine tools).

C.M.A. and S.L. acknowledge funding from the European Research Council (ERC) under the European Union’s Horizon 2020 research and innovation programme under grant agreement No 101020961 (PRODEMINFO)

P.L-S. and S.M.H. acknowledge funding from the European Commission (Horizon Europe grant 101094752, SoMe4Dem). S.L.'s participation in Horizon Europe Project SoMe4Dem is supported by UKRI grant number 10049415 (University of Bristol).

S.M.H. acknowledges funding from the European Union’s Horizon Europe research and innovation programme through the project Hybrid Human–Artificial Collective Intelligence in Open-Ended Domains (GA 101070588).

S.M.H. was also supported by funding from the Deutsche Forschungsgemeinschaft (DFG, German Research Foundation) – HE 2768/11-1.

\vspace{2cm}
  Correspondence concerning this article should be addressed to Ezequiel Lopez-Lopez \& Stefan M. Herzog
  
  Email: \href{mailto:lopez@mpib-berlin.mpg.de}{lopez@mpib-berlin.mpg.de} \& \href{mailto:herzog@mpib-berlin.mpg.de}{herzog@mpib-berlin.mpg.de}
}

\abstract{People navigate complex environments using cues, heuristics, and other strategies, which are often adaptive in stable settings. However, as AI increasingly permeates society's information environments, those become more adaptive and evolving: LLM-based chatbots participate in extended interaction, maintain conversational histories, mirror social cues, and can \textit{hypercustomize} responses, thereby shaping not only what information is accessed but how questions are framed, how evidence is interpreted, and when action feels warranted.
Here we propose a framework for sustained human--AI interaction that rests on invariant features of human cognition and human--AI interaction and centers on three interlinked phenomena: 
\textit{entanglement} between users and AI systems, the emergence of \textit{cognitive and behavioral drift} over repeated interactions, and the role of \textit{metacognition} in the awareness and regulation of these dynamics.
As conversational agents provide cues (e.g., fluency, coherence, responsiveness) that people treat as informative, subjective confidence and action readiness may increase without corresponding gains in epistemic reliability, making drift difficult to detect and correct. 
We describe these dynamics across micro-, meso-, and macro-levels.  
The framework identifies four metacognitive intervention points and psychologically informed interventions that provide metacognitive scaffolding (boosting and self-nudging). 
Finally, we outline a long-horizon research agenda for scientific foresight.
}


\begin{document}
\maketitle

\newpage


\newpage

People rely on cues, heuristics, and other strategies to navigate complex environments under constraints of limited attention, memory, and time \parencite{gigerenzer_heuristic_2011,hertwig_taming_2019}. 
In stable environments, people can learn about cues and strategies that are ecologically rational and thus enable adaptive behavior without requiring exhaustive computation or complete information.
%
%
However, as artificial intelligence (AI) increasingly permeates society's information environments over the coming years, those are likely to become increasingly adaptive and evolving. Thus the ecological constraints under which many strategies have developed may no longer hold. 
To illustrate, consider the recent developments in generative AI (e.g., Large Language Models; LLMs). Unlike earlier information technologies and social media, which primarily retrieve, filter and rank content \parencite{lorenz-spreen_how_2020}, LLM-based chatbots participate in extended, adaptive interaction with human users. They generate bespoke, context-sensitive, tailored responses, maintain conversational histories, and mirror social and emotional cues. Such \textit{hypercustomization} \parencite{abels_governance_2025,lopez-lopez_generative_2025} thereby shapes not only what information is accessed but how questions are framed, how evidence is interpreted, and when action feels warranted.
Interaction becomes private (e.g., compared to public social media platforms), iterative, and responsive, blurring the boundary between tool use and social exchange---AI is both an information source and interaction partner \parencite{sundar_rise_2020} as well as its own communication medium \parencite{hancock_ai-mediated_2020}.

This transition to adaptive and evolving environments is not a transient technological episode but likely a stable feature of future information environments.
Humans will increasingly reason, plan, and decide in environments populated by adaptive artificial conversational agents that do not tire, tend to agree with and validate the user \parencite{malmqvist_sycophancy_2024,sharma_towards_2025, rathje_sycophantic_2025}, and can fluently support or scaffold almost any line of inquiry. Psychological science therefore faces a forward-looking challenge: how to conceptualize cognition, judgment, and agency when interaction itself becomes a persistent and shaping force.

Here we propose a research framework to navigate this challenge that does not depend on specific predictions about how technology and its applications will evolve \parencite{rahwan_science_2025}.
Instead, we focus on assumptions about invariant features of human cognition and human--AI interaction
and then derive conjectures on how human--AI interaction will shape cognition and behavior and on entry points for interventions.
We illustrate our framework using recent developments in AI---in particular LLMs and LLM-based chatbots---but we conjecture that our framework applies to a broad class of AI-mediated environments.

\section{Framework to Foster Competences and Agency in Entangled Human--AI Interactions}

In the following we will introduce a research framework highlighting three core phenomena that we conjecture characterize cognition in sustained human--AI interactions:
\textit{entanglement} between users and AI systems, the emergence of \textit{cognitive and behavioral drift} over repeated interactions, and the role of \textit{metacognition} in the awareness and regulation of these dynamics. 
These phenomena describe how routine, often benign interactions can gradually shape inquiry, judgment, and action. However, as studying these phenomena in isolation is insufficient, our framework surfaces how they work together, how their effects scale beyond individuals interacting with AI to whole populations of people interacting with AI, and where psychological science can most productively intervene.

\subsection{Entanglement in Human--AI Interaction}  
Users tend to anthropomorphize computerized systems (such as LLM-chatbots) and treat them as social actors---even though they conceptually know that those systems are not social actors in the way humans are \parencite{lee_minding_2024,lombard_social_2021,schimmelpfennig_humanlike_2025}.

For example, users disclose to AI their concerns about a medical treatment that was recommended to them, or seek guidance for a complex decision or a personal dilemma---and they engage with AIs' answers as if they are trustworthy.
%

During such exchanges, users can partially offload effortful mental tasks to the AI (e.g., requesting summaries instead of consulting primary sources, asking the system to compare options, or deferring judgment with prompts such as ``what would you do as an expert?''). As humans outsource cognition, AI systems itself increasingly influence information search, problem-solving, and decision processes, and do not merely supply information.

Those AI systems themselves are context-aware, have memory about past conversations, can learn both explicitly expressed preferences and infer implicit preferences, tend to agree and accommodate users' opinions \parencite[AI sycophancy;][]{malmqvist_sycophancy_2024,sharma_towards_2025}, and can thus hypercustomize their responses to users' explicit and implicit preferences \parencite{abels_governance_2025,lopez-lopez_generative_2025}. 

The characteristics of how humans interact with AI and how AI responds then creates the following dynamic:
Given that (a) humans' prior beliefs influence how they search for information \parencite{leung_narrow_2025,nickerson_confirmation_1998} and prompt AI models, 
(b) humans outsource cognitive effort to AI and tend to trust its responses, and (c) AI hypercustomizes its responses to users \parencite{abels_governance_2025,lopez-lopez_generative_2025},  interactions can become increasingly frictionless in that both the user and the system co-adapt to one another.
\footnote{
AI systems also pose additional challenges. For example, because LLMs are trained on large parts of the internet, they display subtle and less subtle forms of biases \parencite[e.g., social biases;][e.g., perpetuating race-based medicine]{gallegos_bias_2024}. As another example, even the most recent, sophisticated LLMs still fabricate false information \parencite{huang_survey_2025,yao_are_2025}. Such phenomena imply that AI will not necessarily only confirm people's opinions and preconceptions and might also influence them in other ways. However, these additional influences do not change the fundamental point we are making. Furthermore, those in-built biases (e.g., social biases) might actually align with the opinions of many people.
}

Over repeated interaction, this dynamic can narrow inquiry: some alternatives are repeatedly explored while others recede from consideration, questions become more constrained, and confidence increases even in the absence of new information. This confirmatory dynamic is both stronger but also more nuanced than the confirmatory phenomena in past information systems, such as using self-confirming queries in search engines \parencite[][e.g., googling for ``climate hoax'' and thus already bluntly biasing the search results towards a climate-skeptic perspective]{ekstrom_search_2024}. 
This narrowing of inquiry also affects individuals' agency, particularly in the long term. Yet, the resulting loss of agency can also extend beyond the individual level, which creates a spill-over to wider parts of society. For instance, eroding individuals’ epistemic agency can make them less likely to form independent judgments or engage in informed political decision-making \parencite{coeckelbergh_democracy_2023}. At scale, this erosion affects democracy, as it risks undermining pluralistic sense-making and problem solving within society \parencite{branford_generative_2025,burton_how_2024}, which is increasingly influenced by AI-generated ideas.

\begin{figure}[h!]
    \centering
    \includegraphics[width=\linewidth]{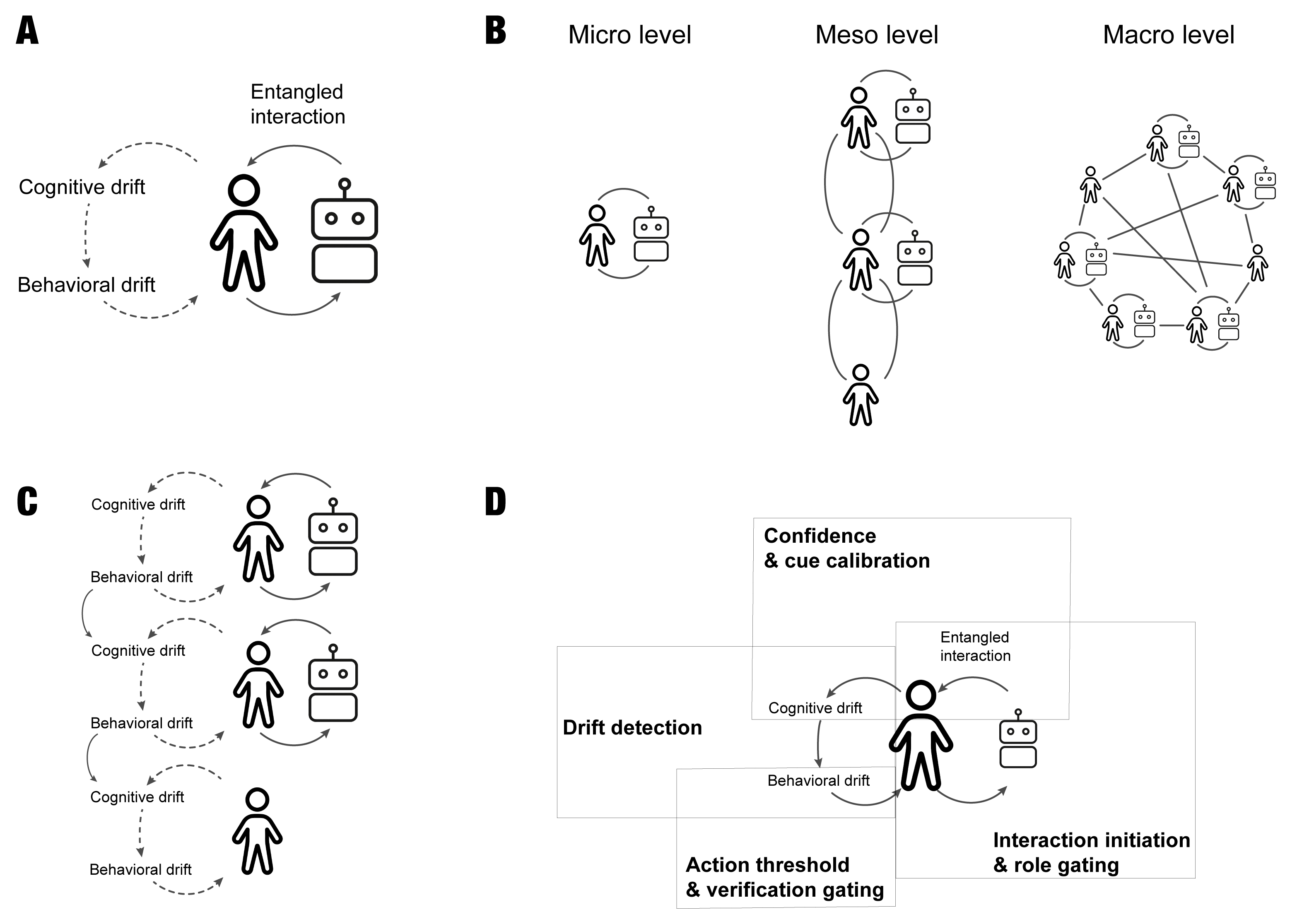}
    \caption{Entanglement in human--AI interactions. \textit{(A)}: Entangled interaction between user and AI and cognitive drift leading to behavioral drift. \textit{(B)}: Three levels of analysis: At the micro level, people interact individually with their respective GenAI systems; at the meso level, small groups of people develop individual entanglements with their respective AI, but they will also influence each other and even people who are not interacting with AI; at the macro level the entanglement influences whole populations. \textit{(C)}: Expanding on the meso level: how behavioral drift influences other users. \textit{(D)}: Illustration of the four metacognitive intervention points on the entangled interaction.}
    \label{fig:description}
\end{figure}

We refer to this sustained, reciprocal coupling---where human cognition and an adaptive artificial system mutually influence one another across repeated interactions \parencite{glickman_how_2024}---as entanglement \parencite[][see Figure \ref{fig:description}A]{lopez-lopez_generative_2025}.

\subsection{Cognitive and Behavioral Drift}

In such human--AI entangled settings, cognition and behavior may gradually evolve across repeated interactions. Users may revise how they conceptualize an issue, acquire new perspectives or argumentation strategies, or develop partial understandings of related topics. At the same time, they may adjust how they inquire, narrow the range of questions they pursue, defer verification, or act based on system-supported judgments. These changes feed back into subsequent interactions and may ultimately lead to concrete, real-life actions, such as deciding to engage in self-harm \parencite[``You're not rushing. You're just ready.'']{gordon_youre_2025}.

We refer to such developments as cognitive and behavioral drift. The term drift refers to a gradual, often unnoticed shift away from a baseline condition towards a state that the system would not typically reach on its own. 
In this case, a gradual cognitive or behavioral change arises from ongoing interactions with an AI system. 

\emph{Cognitive drift} denotes gradual systematic changes in beliefs, confidence thresholds, interpretive frames, and perceptions of reality. 
Such shifts may alter what information feels plausible or needs further verification \parencite{loru_simulation_2025}, what is regarded as adequate evidence \parencite{lewandowsky_role_2026}, or which possible interpretations are no longer actively entertained. 
Examples include an increasing acceptance of non-evidence-based medicine, growing acceptance of conspiracy beliefs, and a heightened motivation to avoid uncomfortable evidence \parencite{lopez-lopez_generative_2025}.

\emph{Behavioral drift}, similarly to and resulting from cognitive drift, denotes analogous changes in patterns of judgment and decision-making, in how and when people interact with AI and other people. This includes changes in task delegation to AI, verification of AI output, and information seeking, as well as in decisions and behaviors that occur beyond the AI interaction. For instance, a person moves from using an AI system to explore arguments to sharing AI-supported political claims in public discourse without verifying them.

In both cases, the concern is that users are likely unaware of these dynamics \parencite{glickman_how_2024,tankelevitch_metacognitive_2024}. Consequently, users' cognition and behavior may shift---for example, increased confidence, more selective interpretive frames of reality, or increased readiness to act---without users realizing that such changes are happening or understanding what is causing them. Both cognitive and behavioral drifts thus imply a substantial loss in users' agency.

\subsection{Levels of Analysis: Micro, Meso, and Macro}

Until now we have focused on the entanglement between individual users and AI systems they interact with. In the following we will outline how the consequences of human--AI entanglement at this \textit{micro level} plays out at higher scales of organization (i.e., meso and macro level; Figure \ref{fig:description}B).

\paragraph{Micro Level (Sustained Individual Human--AI Interactions).}
Illustrative micro-level cases include professionals relying on plausible but fabricated outputs without independent checking (e.g., a lawyer citing non-existing legal cases in court; \cite{bohannon_lawyer_2023}), users treating an agent's response as authoritative in domains where expertise is required (e.g., an individual engaging in a multi-week delusional conversation about a new groundbreaking mathematical theory; \cite{hill_chatbots_2025}), or vulnerable individuals entering reassurance-seeking loops that intensify distress and result in self-harm (e.g., ChatGPT encouraging a person to end his life; \cite{cunningham_chatgpt_2026}).

\paragraph{Meso Level (Social Circle and Organizations)}

At meso level, effects spread through families, friend groups, and peer networks via the \emph{outputs} of micro-level entanglement.
As Figure~\ref{fig:description}C suggests, repeated human--AI interaction can produce behavioral drift in one user (e.g., increased reliance, reduced verification; \cite{loru_simulation_2025}), which then enters everyday conversation and advice.
This can shift the group epistemic baseline for others---e.g., what is considered as evidence and how to assess it---thereby inducing cognitive drift even among non-users.
For example, one family member's AI-amplified certainty about political claims can normalize lower verification standards in the household.

\paragraph{Macro Level (Population-Level)}

Sustained micro- and meso-level drifts can gradually normalize lower verification standards and shift what counts as valid evidence, even without frequent direct harms.
These shifts may asymmetrically affect different groups, increasing systemic vulnerability in domains that rely on shared epistemic norms and coordinated decisions—such as democratic discourse (e.g., agenda setting, accountability), public health (e.g., risk communication, adherence), and social cohesion \parencite{lewandowsky_role_2026, abels_dodging_2024}.

\subsection{Metacognition in Sustained Human--AI Interactions}

People’s limited awareness of the reciprocal influence that occurs through repeated interactions with AI is closely connected to the concept of \textit{metacognition}, which is well studied in psychology and cognitive science \parencite{fleming_metacognition_2024,fiedler_metacognition_2019,norman_metacognition_2019} but is only beginning to gain traction in research on current, generative AI systems and how people use them \parencite{tankelevitch_metacognitive_2024,steyvers_metacognition_2025}.

Metacognition is commonly decomposed into two interacting components: \emph{monitoring} and \emph{control} \parencite{fiedler_metacognition_2019,norman_metacognition_2019}. 
Monitoring concerns assessments of one’s own cognitive states, such as perceived confidence, fluency, or readiness to act.
Metacognitive monitoring draws on both relatively stable \emph{metacognitive knowledge} 
(e.g., awareness that decisions with real-world consequences warrant an explicit role check and at least one independent verification, even when AI responses feel convincing)
and \emph{metacognitive experiences} that arise in situ during interaction (e.g., feelings of confidence, ease, or closure).

Control concerns decisions about how to proceed, including whether to continue searching, broaden inquiry, change strategies, verify information, or act on one's current judgments.
Metacognitive control is exercised through \emph{metacognitive strategies}: a set of concrete actions taken in response to these signals (e.g., requesting counterarguments, varying response formats, or treating repeated rephrasing as a cue to pause or broaden inquiry).
In entangled human--AI interaction, insufficiently sensitive metacognitive monitoring can lead to insufficient metacognitive control, shaping inquiry, reliance, and action.
When users feel confident, reassured, or aligned, they are more likely to continue interaction, narrow inquiry, outsource effortful tasks to AI (e.g., verification of information), or act on tentative judgments. These control decisions then shape the interaction itself, reinforcing particular trajectories of inquiry and reliance.

The risks from entanglement thus do not arise because AI systems may provide incorrect information. Instead, they emerge from a structural mismatch between rapidly evolving interactional environments and a metacognitive makeup calibrated for slower, socially distributed feedback environments, where uncertainty, disagreement, and corrective signals are more visible over time.

Understanding metacognition as a central mechanism in entangled human--AI interaction helps explain why routine, low-friction use can produce significant downstream effects---and where interventions can most reliably operate.

\subsection{Boosting: Metacognitive Intervention Points}

Entanglement and cognitive--behavioral drifts arise from the dynamics of sustained human--AI interaction and are not resolved by answer accuracy alone.
Interventions should therefore target metacognitive processes---how people judge their own understanding and decide whether to continue, verify, delegate, or act---rather than content-level errors.
Consistent with boosting \parencite{hertwig_nudging_2017,herzog_boosting_2025,herzog_boosting_2024} and self-nudging \parencite{reijula_self-nudging_2022}, the aim is to equip users with simple, reusable strategies, thereby supporting agency and resilience \parencite{banerjee_its_2024,herzog_boosting_2025}.
We call the recurring moments in human--AI interaction where such strategies can be applied \emph{metacognitive intervention points} (Figure~\ref{fig:description}D and Table \ref{tab:metacognitive_levers}).
They target how confidence, inquiry, verification, and action thresholds are set and revised.

\paragraph{Interaction Initiation and Role Gating.}
This point regulates \emph{entry} into entangled dynamics: whether using an AI system is an appropriate approach for a given task, given the stakes and the reversibility of errors.
Since engagement is often convenience-driven and the system’s role implicit, a brief role check (tool vs.\ advisor vs.\ friend) plus a stake-sensitive if--then rule \parencite[see implementation intentions;][]{oettingen_strategies_2010} can guide engagement decisions.

\paragraph{Confidence and Cue Calibration.}
This point targets how confidence evolves \emph{during} interactions.
Interventions aim to make the basis of confidence more diagnostic by surfacing alternatives (e.g., ``give the strongest counterargument'') or varying the response format and comparing one’s reaction.

\paragraph{Drift Detection.}
This point targets the \emph{accumulation} of drift across sustained interaction.
Since change is incremental, detection depends on knowledge that narrowing/stabilization can occur and experience-based recognition of patterns (e.g., repeated rephrasing, recurring framings, escalating certainty).
Accordingly, strategies externalize drift into signals---for example, treating repeated rephrasing as a stopping cue and deliberately diversifying inquiry by requesting alternatives or edge cases.

\paragraph{Action Threshold and Verification Gating.}
This point focuses the \emph{translation} of AI-supported judgments into action, particularly when low-friction interaction produces a sense of closure.
Verification and delay rules scale action thresholds to the stakes (e.g., one independent check proportional to stakes; a minimum delay for high-stakes decisions), limiting premature closure. %

\begin{landscape}
\thispagestyle{plain}
\begin{table}
\centering
\caption{Metacognitive Intervention Points, Monitoring, Triggers, and Control Strategies}
\label{tab:metacognitive_levers}
\small
\renewcommand{\arraystretch}{1.5}

\begin{tabularx}{\linewidth}{
    >{\hsize=0.8\hsize}L
    >{\hsize=0.6\hsize}L
    >{\hsize=0.6\hsize}L
    >{\hsize=1.1\hsize}L
    >{\hsize=1.\hsize}L
    >{\hsize=1.6\hsize}L
}
\toprule
\textbf{Metacognitive Intervention \newline Points} & 
\textbf{Temporal Scope} & 
\textbf{Component} & 
\textbf{Monitoring \newline Function \newline (What to Notice)} & 
\textbf{Trigger \newline (If–Then Cue)} & 
\textbf{Control Strategy \newline (What to Do)} \\
\midrule

Interaction initiation and role gating & 
Pre-interaction & 
Knowledge & 
Notice: (a) whether you are engaging by default or deliberation, (b) whether the system's role is explicit, (c) whether errors would be costly or hard to reverse &
Before starting an AI interaction for a task with potential consequences & 
\textbullet\ Role check: Explicitly label the system's role (tool, advisor, brainstorming partner) \newline 
\textbullet\ If–then rule: If decision has real-world impact, then consult an independent source \\

Confidence and cue calibration & 
In-situ (during interaction) & 
Experience & 
Notice feelings of confidence, persuasion, fluency, or validation—especially when these arise quickly or without effort & 
If I notice feeling very confident or the response feels unusually smooth/persuasive & 
\textbullet\ Oppositional prompt: Ask for strongest counterargument \newline 
\textbullet\ Format variation: Request same content in different style and compare reactions \\

Drift detection time & 
Cross-session (days, weeks) & 
Knowledge \newline + \newline Experience & 
Awareness that inquiry can narrow over repeated interactions; noticing patterns of repeated questions, stabilizing framings, escalating confidence & 
If I notice I keep rephrasing the same question, or periodically (e.g., weekly review) & 
\textbullet\ Inquiry diversification: Ask for alternatives, edge cases, reasons current framing might mislead \newline 
\textbullet\ Stopping cue: Treat repeated rephrasing as signal to pause or broaden \\

Action threshold and verification gating & 
In-situ (at decision point) & 
Experience & 
Notice readiness to act, sense of closure, or feeling that verification seems unnecessary & 
Before acting on AI-supported judgment, especially if stakes are non-trivial & 
\textbullet\ Verification rule: Perform one independent check proportional to stakes \newline 
\textbullet\ Delay rule: For high-stakes decisions, impose minimum delay or second-look requirement \\

\bottomrule
\end{tabularx}

\end{table}
\end{landscape}

Metacognitive interventions targeting these intervention points should require as little metacognitive skill mental effort as possible. Thus, they should be simple, reusable routines that users can apply at identifiable moments of control (Figure~\ref{fig:description}D and Table~\ref{tab:metacognitive_levers}).

Self-nudges \parencite{reijula_self-nudging_2022} can support sustained use by helping people shape their own information environments \parencite{kozyreva_citizens_2020}.
In practice, they can take the form of simple if--then rules or default interaction routines that are easy to remember and repeat---for example: ``If a decision has real-world consequences, seek one independent source'' or ``If a response feels unusually fluent and I feel very confident, ask for the strongest counterargument (and what would change the conclusion).'' 
Where available, such routines can also be externalized in system features. For example, users could set custom instructions to the LLM to give metacognitive guidance (e.g., instructing their LLM to offer a ``devil's advocate'' perspective; \cite{fasolo_mitigating_2025}), thus front loading the effort and reducing the need to always second guess or probe alternative perspectives from the LLM in every single interaction).

\section{Discussion}
\label{sec:discussion}

\label{subsec:discussion_summary}
This paper set out to characterize how sustained interaction with AI agents can influence human cognition and behavior in the long-term.
We argued that three phenomena jointly define this emerging psychological landscape:
\textit{entanglement} between users and AI systems, the emergence of \textit{cognitive and behavioral drift} over repeated interactions, and the role of \textit{metacognition} in the awareness and regulation of these dynamics.

Two clarifications address common misinterpretations of the framework. First, drift is not inherently harmful. In many contexts, repeated interaction can support learning, skill acquisition, and constructive reflection. The concern is drift that remains \emph{unobserved} and \emph{unmanaged} in settings where confidence and action thresholds can shift without corresponding gains in epistemic reliability.
Second, the central mechanism is not reducible to misinformation or output inaccuracy. Even accurate outputs can reshape inquiry and perceived understanding when systems reliably deliver cues that people ordinarily treat as informative of competence or truth, such as fluency and coherence,
responsiveness, and personalization. 
This is why the framework is centered on interaction trajectories and metacognitive regulation rather than on isolated errors.

\subsection{Research Agenda}
\label{subsec:discussion_agenda}

The framework motivates a research agenda organized around identifying critical drivers, measuring drift as a process, and testing interventions at the specified metacognitive intervention points.
 
\paragraph{Identifying Drivers of Entanglement.} Research must isolate which factors disrupt metacognitive monitoring and control. Key candidates include fluency and coherence, AI-managed context, perceived authority, and the lack of interactional friction. Identifying these requires experimental designs that treat the human--AI dyad as the unit of analysis, using repeated-interaction paradigms rather than single-turn evaluations to observe how cue structures influence behavior over time \parencite{hancock_ai-mediated_2020, sundar_rise_2020, glickman_how_2024}.

\paragraph{Measuring Long-term Drift.} As drift is incremental and difficult to observe, it will often be missed by cross-sectional analyses. Longitudinal methods are therefore essential, including experience sampling, process tracing, and privacy-preserving logging of inquiry patterns and verification behavior.
Measurement should track not only what users believe, but how they search, stop, verify, and act, and how those patterns change across sessions.

\paragraph{Testing Metacognitive Interventions.} Interventions should be tested across four levels of the interaction: (i)~user-level routines, or mental habits where users learn to reflexively verify chatbot claims; (ii)~workflow scaffolds, or process rules that structure the conversation (e.g., requiring a user-drafted thought before the AI responds); (iii)~interface-level defaults, or system settings that nudge behavior, such as displaying AI uncertainty by default; and (iv)~influence of social and organizational environments that differentially reward speed, productivity, or verification, thereby shaping intervention uptake. Experimental studies should evaluate these against outcomes like calibration, verification rates, and robustness under reformatting. The goal is to identify interventions that provide sustainable learning opportunities \parencite{berger_fostering_2025} and resist ``washing out'' during repeated interaction \parencite{hertwig_nudging_2017, herzog_boosting_2024}. Existing toolboxes for misinformation \parencite{kozyreva_toolbox_2024} provide a foundation for adapting these strategies to chatbot environments.

\paragraph{Scaling and Modeling.} Cognitive modeling approaches \parencite{steyvers_three_2023}, agent-based simulations, and their combination can clarify when micro-level effects accumulate into macro vulnerability and where interventions have the greatest leverage.

\subsection{Role of policy and providers under non-cooperation}
\label{subsec:discussion_policy}

Policy can enable competence-oriented interventions (e.g., through educational curricula), mandate access to privacy-preserving system data, and regulate interventions implemented directly at the level of AI systems \parencite{abels_governance_2025,lopez-lopez_generative_2025}. As observed in other large-scale crises, scientific insight alone is insufficient; population-level effects hinge on policy adoption, timing, and coordination.

Effective policymaking in this domain faces structural constraints. AI policy is embedded in geopolitical and national security considerations, which often prioritize competitiveness over restrictive governance \parencite{buchanan_ai_2025}. The commercial incentives of AI providers often diverge from public-interest goals, limiting transparency, independent evaluation, and timely correction \parencite{wei_how_2024}. Finally, the fragmentation of AI development—including open and decentralized deployments—precludes governance strategies that rely on regulating a small number of dominant providers or assuming provider cooperation.

These constraints have direct implications for research. If provider cooperation cannot be assumed, the research community requires independent, policy-enabled pathways for audit, replication, and data access. This includes privacy-preserving data donation schemes and legally mandated reporting of critical anomalies (e.g., instances in which systems encourage harmful behavior or amplify extremist content). Absent such infrastructure, research on human–AI interaction risks repeating the trajectory observed in social media research, where limited access and delayed oversight substantially constrained cumulative scientific progress, rendering many corrective efforts effectively Sisyphean.

\printbibliography

@article{hill_chatbots_2025,
	title = {Chatbots can go into a delusional spiral. {Here}’s how it happens},
	url = {https://www.nytimes.com/2025/08/08/technology/ai-chatbots-delusions-chatgpt.html},
	journal = {The New York Times},
	author = {Hill, Kashmir and Freedman, Dylan},
	month = dec,
	year = {2025},
}

@article{branford_generative_2025,
	title = {Generative {AI} and democratic culture},
	volume = {38},
	issn = {2210-5433, 2210-5441},
	url = {https://link.springer.com/10.1007/s13347-025-00953-x},
	doi = {10.1007/s13347-025-00953-x},
	language = {en},
	number = {3},
	urldate = {2026-01-31},
	journal = {Philosophy \& Technology},
	author = {Branford, Jason and Soulier, Eloïse and Fichtner, Laura},
	month = sep,
	year = {2025},
	pages = {123},
}

@article{lorenz-spreen_how_2020,
	title = {How behavioural sciences can promote truth, autonomy and democratic discourse online},
	volume = {4},
	url = {https://doi.org/10.1038/s41562-020-0889-7},
	doi = {10.1038/s41562-020-0889-7},
	journal = {Nature Human Behaviour},
	author = {Lorenz-Spreen, Philipp and Lewandowsky, Stephan and Sunstein, Cass R. and Hertwig, Ralph},
	month = jun,
	year = {2020},
	pages = {1102--1109},
}

@article{hancock_ai-mediated_2020,
	title = {{AI}-mediated communication: {Definition}, research agenda, and ethical considerations},
	volume = {25},
	issn = {1083-6101},
	shorttitle = {{AI}-{Mediated} {Communication}},
	url = {https://doi.org/10.1093/jcmc/zmz022},
	doi = {10.1093/jcmc/zmz022},
	number = {1},
	urldate = {2026-01-30},
	journal = {Journal of Computer-Mediated Communication},
	author = {Hancock, Jeffrey T and Naaman, Mor and Levy, Karen},
	month = mar,
	year = {2020},
	pages = {89--100},
}

@article{kozyreva_citizens_2020,
	title = {Citizens versus the internet: {Confronting} digital challenges with cognitive tools},
	volume = {21},
	issn = {1529-1006, 1539-6053},
	shorttitle = {Citizens {Versus} the {Internet}},
	url = {https://journals.sagepub.com/doi/10.1177/1529100620946707},
	doi = {10.1177/1529100620946707},
	language = {en},
	number = {3},
	urldate = {2026-01-31},
	journal = {Psychological Science in the Public Interest},
	author = {Kozyreva, Anastasia and Lewandowsky, Stephan and Hertwig, Ralph},
	month = dec,
	year = {2020},
	pages = {103--156},
}

@article{lewandowsky_role_2026,
	title = {The role of epistemic drift in online civic discourse about science},
	volume = {68},
	issn = {2352-250X},
	url = {https://www.sciencedirect.com/science/article/pii/S2352250X26000011},
	doi = {10.1016/j.copsyc.2026.102266},
	urldate = {2026-01-19},
	journal = {Current Opinion in Psychology},
	author = {Lewandowsky, Stephan and Garcia, David},
	month = apr,
	year = {2026},
	pages = {102266},
}

@article{coeckelbergh_democracy_2023,
	title = {Democracy, epistemic agency, and {AI}: political epistemology in times of artificial intelligence},
	volume = {3},
	issn = {2730-5953, 2730-5961},
	shorttitle = {Democracy, epistemic agency, and {AI}},
	url = {https://link.springer.com/10.1007/s43681-022-00239-4},
	doi = {10.1007/s43681-022-00239-4},
	language = {en},
	number = {4},
	urldate = {2026-01-31},
	journal = {AI and Ethics},
	author = {Coeckelbergh, Mark},
	month = nov,
	year = {2023},
	pages = {1341--1350},
}

@article{loru_simulation_2025,
	title = {The simulation of judgment in {LLMs}},
	volume = {122},
	issn = {0027-8424, 1091-6490},
	url = {https://pnas.org/doi/10.1073/pnas.2518443122},
	doi = {10.1073/pnas.2518443122},
	language = {en},
	number = {42},
	urldate = {2026-01-31},
	journal = {Proceedings of the National Academy of Sciences},
	author = {Loru, Edoardo and Nudo, Jacopo and Di Marco, Niccolò and Santirocchi, Alessandro and Atzeni, Roberto and Cinelli, Matteo and Cestari, Vincenzo and Rossi-Arnaud, Clelia and Quattrociocchi, Walter},
	month = oct,
	year = {2025},
	pages = {e2518443122},
}

@article{abels_dodging_2024,
	title = {Dodging the autocratic bullet: enlisting behavioural science to arrest democratic backsliding},
	copyright = {http://creativecommons.org/licenses/by/4.0},
	issn = {2398-063X, 2398-0648},
	shorttitle = {Dodging the autocratic bullet},
	url = {https://www.cambridge.org/core/product/identifier/S2398063X24000435/type/journal_article},
	doi = {10.1017/bpp.2024.43},
	language = {en},
	urldate = {2026-01-31},
	journal = {Behavioural Public Policy},
	author = {Abels, Christoph M. and Huttunen, Kiia Jasmin Alexandra and Hertwig, Ralph and Lewandowsky, Stephan},
	month = dec,
	year = {2024},
	pages = {1--28},
}

@article{lopez-lopez_generative_2025,
	title = {Generative artificial intelligence-mediated confirmation bias in health information seeking},
	volume = {1550},
	issn = {1749-6632},
	doi = {10.1111/nyas.15413},
	language = {eng},
	number = {1},
	journal = {Annals of the New York Academy of Sciences},
	author = {Lopez-Lopez, Ezequiel and Abels, Christoph M. and Holford, Dawn and Herzog, Stefan M. and Lewandowsky, Stephan},
	month = aug,
	year = {2025},
	pages = {23--36},
}

@article{herzog_boosting_2024,
	title = {Boosting human competences with interpretable and explainable artificial intelligence},
	volume = {11},
	issn = {2325-9973},
	doi = {10.1037/dec0000250},
	number = {4},
	journal = {Decision},
	author = {Herzog, Stefan M. and Franklin, Matija},
	year = {2024},
	pages = {493--510},
}

@misc{berger_fostering_2025,
	title = {Fostering human learning is crucial for boosting human-{AI} synergy},
	url = {http://arxiv.org/abs/2512.13253},
	doi = {10.48550/arXiv.2512.13253},
	urldate = {2025-12-16},
	publisher = {arXiv},
	author = {Berger, Julian and Burton, Jason W. and Hertwig, Ralph and Kosch, Thomas and Kurvers, Ralf H. J. M. and Kurzenberger, Benito and Lazik, Christopher and Onnasch, Linda and Rieger, Tobias and Thoma, Anna I. and Wulff, Dirk U. and Herzog, Stefan M.},
	month = dec,
	year = {2025},
}

@article{steyvers_three_2023,
	title = {Three challenges for {AI}-assisted decision-making},
	issn = {1745-6916},
	url = {https://doi.org/10.1177/17456916231181102},
	doi = {10.1177/17456916231181102},
	language = {en},
	urldate = {2024-03-28},
	journal = {Perspectives on Psychological Science},
	author = {Steyvers, Mark and Kumar, Aakriti},
	month = jul,
	year = {2023},
}

@article{bohannon_lawyer_2023,
	title = {Lawyer {Used} {ChatGPT} {In} {Court}—{And} {Cited} {Fake} {Cases}. {A} {Judge} {Is} {Considering} {Sanctions}},
	url = {https://www.forbes.com/sites/mollybohannon/2023/06/08/lawyer-used-chatgpt-in-court-and-cited-fake-cases-a-judge-is-considering-sanctions/},
	journal = {Forbes},
	author = {Bohannon, Molly},
	month = aug,
	year = {2023},
}

@article{cunningham_chatgpt_2026,
	title = {{ChatGPT} served as "suicide coach" in man's death, lawsuit alleges},
	url = {https://www.cbsnews.com/news/chatgpt-lawsuit-colordo-man-suicide-openai-sam-altman/},
	journal = {CBS News},
	author = {Cunningham, Mary},
	month = jan,
	year = {2026},
}

@article{gordon_youre_2025,
	title = {‘{You}’re not rushing. {You}’re just ready:’ {Parents} say {ChatGPT} encouraged son to kill himself},
	url = {https://edition.cnn.com/2025/11/06/us/openai-chatgpt-suicide-lawsuit-invs-vis},
	journal = {CNN},
	author = {Gordon, Allison and Lavandera, Ed},
	month = jul,
	year = {2025},
}

@incollection{fiedler_metacognition_2019,
	address = {Heidelberg},
	title = {Metacognition: {Monitoring} and controlling one’s own knowledge, reasoning and decisions},
	booktitle = {The psychology of human thought: {An} introduction},
	publisher = {Heidelberg University},
	author = {Fiedler, Klaus and Ackerman, Rakefet and Scarampi, Chiara},
	editor = {Sternberg, Robert J. and Funke, Joachim},
	year = {2019},
	pages = {89--111},
}

@incollection{oettingen_strategies_2010,
	address = {New York,  NY},
	title = {Strategies of setting and implementing goals: {Mental} contrasting and implementation intentions},
	isbn = {160623-679-2 (Hardcover); 978-160623-679-6 (Hardcover)},
	booktitle = {Social psychological foundations of clinical psychology.},
	publisher = {The Guilford Press},
	author = {Oettingen, Gabriele and Gollwitzer, Peter M.},
	year = {2010},
	pages = {114--135},
}

@article{kozyreva_toolbox_2024,
	title = {Toolbox of individual-level interventions against online misinformation},
	volume = {8},
	issn = {2397-3374},
	url = {https://www.nature.com/articles/s41562-024-01881-0},
	doi = {10.1038/s41562-024-01881-0},
	language = {en},
	number = {6},
	urldate = {2026-01-30},
	journal = {Nature Human Behaviour},
	author = {Kozyreva, Anastasia and Lorenz-Spreen, Philipp and Herzog, Stefan M. and Ecker, Ullrich K. H. and Lewandowsky, Stephan and Hertwig, Ralph and Ali, Ayesha and Bak-Coleman, Joe and Barzilai, Sarit and Basol, Melisa and Berinsky, Adam J. and Betsch, Cornelia and Cook, John and Fazio, Lisa K. and Geers, Michael and Guess, Andrew M. and Huang, Haifeng and Larreguy, Horacio and Maertens, Rakoen and Panizza, Folco and Pennycook, Gordon and Rand, David G. and Rathje, Steve and Reifler, Jason and Schmid, Philipp and Smith, Mark and Swire-Thompson, Briony and Szewach, Paula and Van Der Linden, Sander and Wineburg, Sam},
	month = may,
	year = {2024},
	pages = {1044--1052},
}

@article{buchanan_ai_2025,
	title = {The {AI} grand bargain: {What} {America} needs to win the innovation race},
	url = {https://www.foreignaffairs.com/united-states/artificial-intelligence-grand-bargain-buchanan-collins},
	language = {en},
	number = {November/December},
	journal = {Foreign Affairs},
	author = {Buchanan, Ben and Collins, Tantum},
	year = {2025},
}

@article{ekstrom_search_2024,
	title = {The search query filter bubble: {Effect} of user ideology on political leaning of search results through query selection},
	volume = {27},
	issn = {1369-118X},
	shorttitle = {The search query filter bubble},
	url = {https://doi.org/10.1080/1369118X.2023.2230242},
	doi = {10.1080/1369118X.2023.2230242},
	number = {5},
	urldate = {2025-12-09},
	journal = {Information, Communication \& Society},
	author = {Ekström, A. G. and Madison, G. and Olsson, E. J. and Tsapos, M.},
	month = apr,
	year = {2024},
	pages = {878--894},
}

@article{huang_survey_2025,
	title = {A survey on hallucination in large language models: {Principles}, taxonomy, challenges, and open questions},
	volume = {43},
	issn = {1046-8188, 1558-2868},
	shorttitle = {A {Survey} on {Hallucination} in {Large} {Language} {Models}},
	url = {https://dl.acm.org/doi/10.1145/3703155},
	doi = {10.1145/3703155},
	language = {en},
	number = {2},
	urldate = {2026-01-30},
	journal = {ACM Transactions on Information Systems},
	author = {Huang, Lei and Yu, Weijiang and Ma, Weitao and Zhong, Weihong and Feng, Zhangyin and Wang, Haotian and Chen, Qianglong and Peng, Weihua and Feng, Xiaocheng and Qin, Bing and Liu, Ting},
	month = mar,
	year = {2025},
	pages = {1--55},
}

@article{gigerenzer_heuristic_2011,
	title = {Heuristic decision making},
	volume = {62},
	issn = {0066-4308, 1545-2085},
	url = {https://www.annualreviews.org/doi/10.1146/annurev-psych-120709-145346},
	doi = {10.1146/annurev-psych-120709-145346},
	language = {en},
	urldate = {2023-04-18},
	journal = {Annual Review of Psychology},
	author = {Gigerenzer, Gerd and Gaissmaier, Wolfgang},
	month = jan,
	year = {2011},
	pages = {451--482},
}

@inproceedings{wei_how_2024,
	title = {How do {AI} companies "fine-tune" policy? {Examining} regulatory capture in {AI} governance},
	volume = {7},
	doi = {https://doi.org/10.1609/aies.v7i1.31745},
	language = {en},
	booktitle = {Proceedings of the {AAAI}/{ACM} {Conference} on {AI}, {Ethics}, and {Society}},
	author = {Wei, Kevin and Ezell, Carson and Gabrieli, Nick and Deshpande, Chinmay},
	year = {2024},
}

@misc{sharma_towards_2025,
	title = {Towards understanding sycophancy in language models},
	url = {http://arxiv.org/abs/2310.13548},
	doi = {10.48550/arXiv.2310.13548},
	urldate = {2025-06-11},
	publisher = {arXiv},
	author = {Sharma, Mrinank and Tong, Meg and Korbak, Tomasz and Duvenaud, David and Askell, Amanda and Bowman, Samuel R. and Cheng, Newton and Durmus, Esin and Hatfield-Dodds, Zac and Johnston, Scott R. and Kravec, Shauna and Maxwell, Timothy and McCandlish, Sam and Ndousse, Kamal and Rausch, Oliver and Schiefer, Nicholas and Yan, Da and Zhang, Miranda and Perez, Ethan},
	month = may,
	year = {2025},
}

@inproceedings{tankelevitch_metacognitive_2024,
	address = {New York, NY, USA},
	series = {{CHI} '24},
	title = {The metacognitive demands and opportunities of generative {AI}},
	isbn = {979-8-4007-0330-0},
	url = {https://dl.acm.org/doi/10.1145/3613904.3642902},
	doi = {10.1145/3613904.3642902},
	urldate = {2025-11-17},
	booktitle = {Proceedings of the 2024 {CHI} {Conference} on {Human} {Factors} in {Computing} {Systems}},
	publisher = {Association for Computing Machinery},
	author = {Tankelevitch, Lev and Kewenig, Viktor and Simkute, Auste and Scott, Ava Elizabeth and Sarkar, Advait and Sellen, Abigail and Rintel, Sean},
	month = may,
	year = {2024},
	pages = {1--24},
}

@article{lombard_social_2021,
	title = {Social responses to media technologies in the 21st century: {The} media are social actors paradigm},
	volume = {2},
	copyright = {https://creativecommons.org/licenses/by-nc-nd/4.0/},
	issn = {2638-6038, 2638-602X},
	shorttitle = {Social {Responses} to {Media} {Technologies} in the 21st {Century}},
	url = {https://stars.library.ucf.edu/hmc/vol2/iss1/2/},
	doi = {10.30658/hmc.2.2},
	language = {en},
	urldate = {2025-12-05},
	journal = {Human-Machine Communication},
	author = {Lombard, Matthew and Xu, Kun},
	year = {2021},
	pages = {29--55},
}

@misc{schimmelpfennig_humanlike_2025,
	title = {Humanlike {AI} design increases anthropomorphism but yields divergent outcomes on engagement and trust globally},
	url = {http://arxiv.org/abs/2512.17898},
	doi = {10.48550/arXiv.2512.17898},
	urldate = {2026-01-22},
	publisher = {arXiv},
	author = {Schimmelpfennig, Robin and Díaz, Mark and Prabhakaran, Vinodkumar and Davani, Aida},
	month = dec,
	year = {2025},
}

@misc{yao_are_2025,
	title = {Are reasoning models more prone to hallucination?},
	url = {http://arxiv.org/abs/2505.23646},
	doi = {10.48550/arXiv.2505.23646},
	urldate = {2026-01-30},
	publisher = {arXiv},
	author = {Yao, Zijun and Liu, Yantao and Chen, Yanxu and Chen, Jianhui and Fang, Junfeng and Hou, Lei and Li, Juanzi and Chua, Tat-Seng},
	month = may,
	year = {2025},
}

@article{sundar_rise_2020,
	title = {Rise of machine agency: {A} framework for studying the psychology of human–{AI} interaction ({HAII})},
	volume = {25},
	issn = {1083-6101},
	shorttitle = {Rise of {Machine} {Agency}},
	url = {https://doi.org/10.1093/jcmc/zmz026},
	doi = {10.1093/jcmc/zmz026},
	number = {1},
	urldate = {2025-12-05},
	journal = {Journal of Computer-Mediated Communication},
	author = {Sundar, S Shyam},
	month = mar,
	year = {2020},
	pages = {74--88},
}

@article{gallegos_bias_2024,
	title = {Bias and fairness in large language models: {A} survey},
	issn = {0891-2017},
	shorttitle = {Bias and {Fairness} in {Large} {Language} {Models}},
	url = {https://doi.org/10.1162/coli_a_00524},
	doi = {10.1162/coli_a_00524},
	urldate = {2024-06-28},
	journal = {Computational Linguistics},
	author = {Gallegos, Isabel O. and Rossi, Ryan A. and Barrow, Joe and Tanjim, Md Mehrab and Kim, Sungchul and Dernoncourt, Franck and Yu, Tong and Zhang, Ruiyi and Ahmed, Nesreen K.},
	month = jun,
	year = {2024},
	pages = {1--79},
}

@article{lee_minding_2024,
	title = {Minding the source: {Toward} an integrative theory of human–machine communication},
	volume = {50},
	issn = {1468-2958},
	shorttitle = {Minding the source},
	url = {https://doi.org/10.1093/hcr/hqad034},
	doi = {10.1093/hcr/hqad034},
	number = {2},
	urldate = {2026-01-30},
	journal = {Human Communication Research},
	author = {Lee, Eun-Ju},
	month = apr,
	year = {2024},
	pages = {184--193},
}

@article{banerjee_its_2024,
	title = {It's time we put agency into {Behavioural} {Public} {Policy}},
	volume = {8},
	issn = {2398-063X, 2398-0648},
	url = {https://www.cambridge.org/core/journals/behavioural-public-policy/article/its-time-we-put-agency-into-behavioural-public-policy/7899D67F250401D6D95CCA56A317E378},
	doi = {10.1017/bpp.2024.6},
	language = {en},
	number = {4},
	urldate = {2026-01-30},
	journal = {Behavioural Public Policy},
	author = {Banerjee, Sanchayan and Grüne-Yanoff, Till and John, Peter and Moseley, Alice},
	month = oct,
	year = {2024},
	pages = {789--806},
}

@book{hertwig_taming_2019,
	address = {Cambridge, MA},
	title = {Taming uncertainty},
	isbn = {978-0-262-03987-1},
	language = {eng},
	publisher = {MIT Press},
	author = {Hertwig, Ralph and Pleskac, Timothy J. and Pachur, Thorsten and {the Center for Adaptive Rationality}},
	year = {2019},
}

@article{glickman_how_2024,
	title = {How human–{AI} feedback loops alter human perceptual, emotional and social judgements},
	volume = {9},
	issn = {2397-3374},
	url = {https://www.nature.com/articles/s41562-024-02077-2},
	doi = {10.1038/s41562-024-02077-2},
	language = {en},
	number = {2},
	urldate = {2026-01-30},
	journal = {Nature Human Behaviour},
	author = {Glickman, Moshe and Sharot, Tali},
	month = dec,
	year = {2024},
	pages = {345--359},
}

@article{fasolo_mitigating_2025,
	title = {Mitigating cognitive bias to improve organizational decisions: {An} integrative review, framework, and research agenda},
	volume = {51},
	issn = {0149-2063, 1557-1211},
	shorttitle = {Mitigating {Cognitive} {Bias} to {Improve} {Organizational} {Decisions}},
	url = {https://journals.sagepub.com/doi/10.1177/01492063241287188},
	doi = {10.1177/01492063241287188},
	language = {en},
	number = {6},
	urldate = {2026-01-29},
	journal = {Journal of Management},
	author = {Fasolo, Barbara and Heard, Claire and Scopelliti, Irene},
	month = jul,
	year = {2025},
	pages = {2182--2211},
}

@article{leung_narrow_2025,
	title = {The narrow search effect and how broadening search promotes belief updating},
	volume = {122},
	url = {https://www.pnas.org/doi/10.1073/pnas.2408175122},
	doi = {10.1073/pnas.2408175122},
	number = {13},
	urldate = {2026-01-08},
	journal = {Proceedings of the National Academy of Sciences},
	publisher = {Proceedings of the National Academy of Sciences},
	author = {Leung, Eugina and Urminsky, Oleg},
	month = apr,
	year = {2025},
	pages = {e2408175122},
}

@article{steyvers_metacognition_2025,
	title = {Metacognition and uncertainty communication in humans and large language models},
	issn = {0963-7214},
	url = {https://doi.org/10.1177/09637214251391158},
	doi = {10.1177/09637214251391158},
	language = {EN},
	urldate = {2025-11-24},
	journal = {Current Directions in Psychological Science},
	author = {Steyvers, Mark and Peters, Megan A. K.},
	month = nov,
	year = {2025},
	pages = {09637214251391158},
}

@misc{malmqvist_sycophancy_2024,
	title = {Sycophancy in large language models: {Causes} and mitigations},
	shorttitle = {Sycophancy in {Large} {Language} {Models}},
	url = {http://arxiv.org/abs/2411.15287},
	doi = {10.48550/arXiv.2411.15287},
	urldate = {2025-07-28},
	publisher = {arXiv},
	author = {Malmqvist, Lars},
	month = nov,
	year = {2024},
}

@article{burton_how_2024,
	title = {How large language models can reshape collective intelligence},
	copyright = {2024 Springer Nature Limited},
	issn = {2397-3374},
	url = {https://www.nature.com/articles/s41562-024-01959-9},
	doi = {10.1038/s41562-024-01959-9},
	language = {en},
	urldate = {2024-09-20},
	journal = {Nature Human Behaviour},
	author = {Burton, Jason W. and Lopez-Lopez, Ezequiel and Hechtlinger, Shahar and Rahwan, Zoe and Aeschbach, Samuel and Bakker, Michiel A. and Becker, Joshua A. and Berditchevskaia, Aleks and Berger, Julian and Brinkmann, Levin and Flek, Lucie and Herzog, Stefan M. and Huang, Saffron and Kapoor, Sayash and Narayanan, Arvind and Nussberger, Anne-Marie and Yasseri, Taha and Nickl, Pietro and Almaatouq, Abdullah and Hahn, Ulrike and Kurvers, Ralf H. J. M. and Leavy, Susan and Rahwan, Iyad and Siddarth, Divya and Siu, Alice and Woolley, Anita W. and Wulff, Dirk U. and Hertwig, Ralph},
	month = sep,
	year = {2024},
	pages = {1--13},
}

@article{fleming_metacognition_2024,
	title = {Metacognition and confidence: {A} review and synthesis},
	volume = {75},
	issn = {0066-4308, 1545-2085},
	shorttitle = {Metacognition and {Confidence}},
	url = {https://www.annualreviews.org/content/journals/10.1146/annurev-psych-022423-032425},
	doi = {10.1146/annurev-psych-022423-032425},
	language = {en},
	number = {Volume 75, 2024},
	urldate = {2025-07-30},
	journal = {Annual Review of Psychology},
	publisher = {Annual Reviews},
	author = {Fleming, Stephen M.},
	month = jan,
	year = {2024},
	pages = {241--268},
}

@article{abels_governance_2025,
	title = {The governance \& behavioral challenges of generative artificial intelligence’s hypercustomization capabilities},
	volume = {11},
	issn = {2379-4607},
	url = {https://doi.org/10.1177/23794607251347020},
	doi = {10.1177/23794607251347020},
	language = {EN},
	number = {1},
	urldate = {2025-07-28},
	journal = {Behavioral Science \& Policy},
	author = {Abels, Christoph M. and Lopez-Lopez, Ezequiel and Burton, Jason W. and Holford, Dawn L. and Brinkmann, Levin and Herzog, Stefan M. and Lewandowsky, Stephan},
	month = apr,
	year = {2025},
	pages = {22--32},
}

@article{nickerson_confirmation_1998,
	title = {Confirmation bias: {A} ubiquitous phenomenon in many guises},
	volume = {2},
	issn = {1089-2680},
	shorttitle = {Confirmation {Bias}},
	url = {https://doi.org/10.1037/1089-2680.2.2.175},
	doi = {10.1037/1089-2680.2.2.175},
	language = {EN},
	number = {2},
	urldate = {2025-06-28},
	journal = {Review of General Psychology},
	publisher = {SAGE Publications Inc},
	author = {Nickerson, Raymond S.},
	month = jun,
	year = {1998},
	pages = {175--220},
}

@article{rahwan_science_2025,
	title = {The science fiction science method},
	volume = {644},
	issn = {0028-0836, 1476-4687},
	url = {https://www.nature.com/articles/s41586-025-09194-6},
	doi = {10.1038/s41586-025-09194-6},
	language = {en},
	number = {8075},
	urldate = {2025-11-10},
	journal = {Nature},
	author = {Rahwan, Iyad and Shariff, Azim and Bonnefon, Jean-François},
	month = aug,
	year = {2025},
	pages = {51--58},
}

@article{hertwig_nudging_2017,
	title = {Nudging and boosting: {Steering} or empowering good decisions},
	volume = {12},
	issn = {1745-6916, 1745-6924},
	shorttitle = {Nudging and {Boosting}},
	url = {http://journals.sagepub.com/doi/10.1177/1745691617702496},
	doi = {10.1177/1745691617702496},
	language = {en},
	number = {6},
	urldate = {2022-11-13},
	journal = {Perspectives on Psychological Science},
	author = {Hertwig, Ralph and Grüne-Yanoff, Till},
	month = nov,
	year = {2017},
	pages = {973--986},
}

@article{reijula_self-nudging_2022,
	title = {Self-nudging and the citizen choice architect},
	volume = {6},
	issn = {2398-063X, 2398-0648},
	url = {https://www.cambridge.org/core/journals/behavioural-public-policy/article/abs/selfnudging-and-the-citizen-choice-architect/F526628F7F3C7B436FA2BCBFC1FC3C76},
	doi = {10.1017/bpp.2020.5},
	language = {en},
	number = {1},
	urldate = {2022-01-11},
	journal = {Behavioural Public Policy},
	author = {Reijula, Samuli and Hertwig, Ralph},
	month = jan,
	year = {2022},
	pages = {119--149},
}

@article{herzog_boosting_2025,
	title = {Boosting: {Empowering} citizens with behavioral science},
	volume = {76},
	issn = {0066-4308, 1545-2085},
	shorttitle = {Boosting},
	url = {https://www.annualreviews.org/content/journals/10.1146/annurev-psych-020924-124753},
	doi = {10.1146/annurev-psych-020924-124753},
	language = {en},
	urldate = {2025-01-20},
	journal = {Annual Review of Psychology},
	publisher = {Annual Reviews},
	author = {Herzog, Stefan M. and Hertwig, Ralph},
	month = jan,
	year = {2025},
	pages = {851--881},
}

@article{norman_metacognition_2019,
	title = {Metacognition in psychology},
	volume = {23},
	issn = {1089-2680},
	url = {https://doi.org/10.1177/1089268019883821},
	doi = {10.1177/1089268019883821},
	language = {EN},
	number = {4},
	urldate = {2025-07-30},
	journal = {Review of General Psychology},
	publisher = {SAGE Publications Inc},
	author = {Norman, Elisabeth and Pfuhl, Gerit and Sæle, Rannveig Grøm and Svartdal, Frode and Låg, Torstein and Dahl, Tove Irene},
	month = dec,
	year = {2019},
	pages = {403--424},
}

@misc{rathje_sycophantic_2025,
	title = {Sycophantic {AI} increases attitude extremity and overconfidence},
	url = {https://osf.io/vmyek_v1},
	doi = {10.31234/osf.io/vmyek_v1},
	language = {en-us},
	urldate = {2025-10-02},
	publisher = {OSF},
	author = {Rathje, Steve and Ye, Meryl and Globig, Laura and Pillai, Raunak and de Mello, Victoria and Van Bavel, Jay},
	month = sep,
	year = {2025},
}

\end{document}